\documentclass[journal]{IEEEtran}
\usepackage[caption=false,font=normalsize,labelfont=sf,textfont=sf]{subfig}
\usepackage{cite}
\usepackage{graphics}
\usepackage{amsmath}		
\usepackage{amsfonts}		
\usepackage{amssymb}        
\usepackage{ulem}
\usepackage{svg}
\ifCLASSINFOpdf
\usepackage{graphicx}
\usepackage{txfonts}
\else
\usepackage{graphicx}
\fi
\usepackage{float}
\usepackage{tabu}
\usepackage{booktabs}
\usepackage{placeins}
\usepackage{multicol}
\usepackage{epstopdf}
\usepackage{breqn}
\graphicspath{ {./Images/} }
\usepackage{multirow}
\usepackage{caption}
\usepackage{soul}
\usepackage{algorithm}
\usepackage{algpseudocode}
\usepackage{mathtools}
\usepackage{xcolor}

\usepackage{bbm}
\usepackage{dblfloatfix}

\usepackage{hyperref}
\hypersetup{
    colorlinks=true,
    linkcolor=blue,
    filecolor=magenta,      
    urlcolor=blue,
}

\definecolor{amber(sae/ece)}{rgb}{1.0, 0.49, 0.0}
\definecolor{aqua}{rgb}{0.0, 1.0, 1.0}

\begin{document}

\title{Hardware Software Co-design of Statistical and Deep Learning Frameworks for Wideband Sensing on Zynq System on Chip}

\author{Rohith Rajesh, Sumit J. Darak, Akshay Jain, Shivam Chandhok,  and Animesh Sharma
					\thanks{This work is supported by the funding received from core research grant (CRG) awarded to Dr. Sumit J. Darak from DST-SERB, GoI.}
			\thanks{Rohith Rajesh, Sumit J. Darak and Animesh Sharma are with Electronics and Communications Department, 
				IIIT-Delhi, India-110020 (e-mail: \{rohith18182,sumit,animesh20317\}@iiitd.ac.in}
			\thanks{Akshay Jain is associated with AMD, Hyderabad, India and Electronics and Communications Department, 
				IIIT-Delhi, India. email: akshayj@amd.com}	
				\thanks{Shivam Chandhok is with INRIA, Universite Grenoble Alpes, France}
}
	\maketitle
\begin{abstract}
With the introduction of spectrum sharing and heterogeneous services in next-generation networks, the base stations need to sense the wideband spectrum and identify the spectrum resources to meet the quality-of-service, bandwidth, and latency constraints. Sub-Nyquist sampling (SNS) enables digitization for sparse wideband spectrum without needing Nyquist speed analog-to-digital converters. However, SNS demands additional signal processing algorithms for spectrum reconstruction, such as the well-known orthogonal matching pursuit (OMP) algorithm. OMP is also widely used in other compressed sensing applications. The first contribution of this work is efficiently mapping the OMP algorithm on the Zynq system-on-chip (ZSoC) consisting of an ARM processor and FPGA. Experimental analysis shows a significant degradation in OMP performance for sparse spectrum. Also, OMP needs prior knowledge of spectrum sparsity. We address these challenges via deep-learning-based architectures and efficiently map them on the ZSoC platform as second contribution. Via hardware-software co-design, different versions of the proposed architecture obtained by partitioning between software (ARM processor) and hardware (FPGA) are considered. The resource, power, and execution time comparisons for given memory constraints and a wide range of word lengths are presented for these architectures.

\end{abstract}
\newcommand{\pdob}{$P_d^{OB}$} 
\newcommand{\pdab}{$P_d^{AB}$} 
\section{Introduction}
\label{Sec:Intro}
Wideband spectrum sensing (WSS) involves digitizing the wideband spectrum and identifying available spectrum for resource allocation in wireless networks \cite{WSS1,WSS2,LTE_unlicensed}. WSS has gained significant importance in 5G and next-generation wireless networks due to the introduction of spectrum sharing policies replacing the conventional static spectrum allocation policies \cite{NR1}. Spectrum sharing allows the deployment of the wireless networks in licensed, shared, and unlicensed spectrum \cite{5G1}. The advantages include significant cost savings for Telecom service operators since they can reduce licensed spectrum requirement and corresponding exorbitant spectrum license fee, which is usually in billions of dollars \cite{5G2}. A broader spectrum enables the deployment of upcoming heterogeneous services that demand a wide range of bandwidth, quality-of-service, and latency constraints. Though the millimeter wave spectrum (above 6 GHz) is being explored due to the availability of large bandwidth, various studies have confirmed that it is not a feasible alternative to the sub-6 GHz spectrum for reliable outdoor communication and comprehensive network coverage. 

Numerous spectrum measurement and utilization studies have shown that overall utilization of the sub-6 GHz spectrum is poor even though most of the spectrum is licensed \cite{SM1,SM2,SM3}. For digitization of such a sparse wideband spectrum, sub-Nyquist sampling (SNS) based WSS is an efficient and feasible alternative to Nyquist sampling based WSS. This is because SNS needs low-speed analog-to-digital converters (ADCs) compared to Nyquist rate ADCs in the latter \cite{WSS2,WSS3}. However, SNS-based WSS needs additional digital signal processing to recover the SNS sampled spectrum so that it closely resembles the original spectrum \cite{recovery1,recovery2}. Such recovery must be done accurately and should meet the stringent area, power, and latency constraints. 

Various signal recovery techniques for SNS have been reviewed in \cite{sparsereview}. The greedy-approach-based orthogonal matching pursuit (OMP) framework is popular due to its lower complexity and faster execution time. Other applications of OMP includes image processing, and radar systems \cite{ompapp1,ompapp2,ompapp3,ompapp4}. Recently, advances in deep learning (DL) have been explored to improve the performance of the OMP \cite{Palangi2017ConvolutionalDS,palangi_stacking,7457684}. Most of these approaches augment the OMP by replacing the matched filtering task with the DL. The OMP and its variants suffer from a significant degradation in reconstruction performance, especially when the spectrum sparsity is high. Also, they need prior knowledge of spectrum sparsity which may not be available in a dynamic environment. From an architecture perspective, in 5G and next-generation wireless networks, distributed base station approach where radio unit (RU), distributed unit (DU), and central unit (CU) may not be co-located is being explored \cite{5Garch1,5Garch2}. Such deployment demands the mapping of SNS algorithms on hardware platforms such as system on chip (SoC) under the limited computing resource constraints due to remote locations of the RU and DU. 
	
The first contribution of this work is to efficiently map statistical signal processing based OMP framework for SNS spectrum recovery on Zynq system-on-chip (ZSoC). The ZSoC is the heterogeneous SoC comprising of software (SW), i.e., dual-core Cortex ARM A9 processor, and hardware (HW), i.e., 7-series field programmable gate array (FPGA). To address the drawbacks of OMP  and make the spectrum reconstruction agnostic to sparsity, we replace OMP with a convolutional neural network (CNN) based deep learning (DL) architecture and efficiently map it on the ZSoC platform as the second contribution. Via hardware-software co-design, we explore different versions of the proposed DL architecture obtained by partitioning between SW (ARM processor) and HW (FPGA). Our study offers interesting insights which may not be visible in conventional theoretical and simulation-based analysis. For these architectures, we present the resource utilization, power consumption, and execution time comparisons for given memory constraints and a wide range of word lengths (WL) at the parameters and computation levels. We develop an end-to-end application with a live graphical user interface (GUI) to demonstrate the real-time WSS on the ZSoC platform. 
{\color{black}  Please refer to \cite{codes} for source codes, hardware IPs, datasets and detailed tutorial used for generating the results presented in this paper.}

	The rest of the paper is organized as follows. We discuss the OMP architecture and experimental results in Section~\ref{Sec:omp_analysis}. The DLWSS algorithm is presented in Section~\ref{Sec:DLSWSS_alg} and corresponding architecture in Section~\ref{Sec:DLSWSS_arch}. Section~\ref{Sec:PM} presents the performance analysis results and comparison of DLWSS with OMP followed by complexity analysis in Section~\ref{Sec:CC}. Section~\ref{Sec:conc} concludes the paper. 
	


\section{Spectrum Recovery via OMP for SNS based WSS} \label{Sec:omp_analysis}
This section discusses the design and implementation of the OMP on the ZSoC. We demonstrate the various drawbacks of OMP via experimental analysis and results.



\subsection{Orthogonal Matching Pursuit (OMP) on ZSoC} \label{Sec:omp_hardware}
 OMP \cite{omp_ref,omp_review,omp_proposed_paper} is one of the most widely used sparse recovery algorithm for SNS based WSS. It follows an iterative formulation where an occupied band that is highly correlated with the residual matrix is identified. Then its contribution is removed from the residual matrix to identify the next highly correlated occupied band. This process is repeated for all occupied bands when we have prior knowledge of the spectrum sparsity. Otherwise, we need to have a stopping criterion based on estimated sparsity. The OMP pseudo-code is given in Algorithm~\ref{alg:cap}. Here, $K$ represents the number of ADCs used for SNS, $Q$ is the number of snapshots produced by the ADC, and $N$ is the number of frequency bands in the wideband spectrum. Further, $A$ is the sensing matrix of dimension $K \times N$, $Y$ are the received SNS samples of dimension $N \times Q$, and $\dagger$ represents a matrix pseudo-inverse.
It comprises four main steps, i.e., 1) Matching (Line 6-8); 2) Identification (Line 9-13); 3) Least squares (Line 14) and 4) Approximation (Line 15). 





\begin{algorithm}
\caption{OMP algorithm for WSS}\label{alg:cap}

\begin{algorithmic}[1]
\Require \texttt{Sensing matrix A[K][N], aliased sub-Nyquist samples Y [K] [Q], Sparsity $S$}
\State $\mathtt{occupied\_bands \gets []}$
\State $\mathtt{A_{norm} \gets}$ \texttt{Column normalized A} \Comment{\texttt{Normalize A}}
\State $\mathtt{Res \gets Y}$ \Comment{\texttt{Initialize Residual}}
\State $\mathtt{iter \gets 1}$
\While{$\mathtt{iter \leq S}$} \Comment{\texttt{No. of occupied bands}}
\For{\texttt{j in columns($\mathtt{A_{norm}}$)}}
\State $\mathtt{Z[j] \gets norm(A_{norm}[\;:\;,j\;]^T \times Res)}$
\EndFor
\State \texttt{Append $\mathtt{argmax(Z)\: to\: occupied\_bands}$}
\For{\texttt{j in $\mathtt{occupied\_bands}$}}
\State \texttt{Append column $\mathtt{A_{norm}[\;:\;,j\;]}$ to As}
\EndFor
\State $\mathtt{Res \gets Res - As \times (As^\dagger \times Res)} $
\State $\mathtt{iter\gets iter+1}$
\EndWhile \\
\Return $\mathtt{occupied\_bands}$
\end{algorithmic}
\end{algorithm}

We have realized the OMP in Algorithm~\ref{alg:cap} on ZSoC ZC706 comprising of Dual ARM Cortex A9 processor and 7-series FPGA with 1090 units of 18kB Block RAMs, 80 DSP48E units, 218600 units of 6-input look-up-tables (LUTs) and 437200 flip-flops (FFs).  Table \ref{tab:omp_time} shows the execution time, resource utilization, and power consumption comparison for the two best possible realizations of the OMP algorithm: 1) Only software (ARM), and 2) Software and hardware co-design (ARM+FPGA). Note that we have optimized the code for software implementation and carefully chosen the word length on hardware to minimize the execution time, power consumption, and resource utilization without compromising the functional accuracy. Interestingly, FPGA realization is slower than ARM processors due to the sequential nature of the OMP algorithm and the need for variable size matrix inversion operations. On FPGA, we need to implement single matrix inversion architecture for the largest possible matrix size and use the same for other matrices. Thus, the execution time of matrix inversion operations in all stages is the same on FPGA, irrespective of matrix size.
In contrast, matrix inversion time in SW increases with the increase in matrix size. Even though the OMP is realized on FPGA in the second case, we still need an ARM processor for running the operating system and control operations in remote edge deployment. Thus, the execution of the second realization has a significant contribution from the data communication overhead between ARM and FPGA. Also, the relatively small dimension of matrices means that the ARM processor can execute the arithmetic operations faster due to efficient caching and higher clock frequency than FPGA.


\begin{table}[!t]
\caption{Comparison of execution time, resource and power statistics of OMP realization on ZC706 Platform}
\label{tab:omp_time}
\renewcommand{\arraystretch}{1.1}
 \resizebox{0.5\textwidth}{!}{
\begin{tabular}{@{}|l|l|l|l|@{}}
\hline
\textbf{Configuration} & \textbf{\begin{tabular}[c]{@{}l@{}}Execution Time \\  (Seconds)\end{tabular}} & \textbf{\begin{tabular}[c]{@{}l@{}}Resource Utilization \\ \textbf{\{BRAM,DSP,FF,LUT\}}\end{tabular}}   & \textbf{Total Power (W)} \\ \hline
SW & 0.103 & -- & 1.57 \\ \hline
SW+HW & 0.189 & \{44,258,17160,13070\} & 3.664 \\ \hline
\end{tabular}}
\end{table}

\subsection{Functional Performance Analysis of OMP on ZSoC}\label{section:func_perf_analysis_OMP}
In this section, we analyze the functional correctness of the OMP algorithm realized on ZSoC. To begin with, we discuss the performance metrics used for such analysis. 
\subsubsection{\textbf{Performance Metrics}}
We consider two well-know performance metrics: Detection accuracy of all bands in percentage (\pdab) and detection accuracy of occupied bands in percentage (\pdob) \cite{metric_p1,metric_p2}.
The metric, \pdab, corresponds to the fraction of frequency bands whose status is correctly detected by the algorithm. The second metric, \pdob, corresponds to the fraction of occupied bands whose status is correctly detected by the algorithm. Mathematically, \pdab and \pdob are formulated as:
\begin{center}
    $P_d^{AB}= 100*\frac{\sum_{n=1}^{N} (y_{pred}^n \, == \, y_{true}^n)   }{\sum_{n=1}^{N} (y_{true}^n) }$
\end{center}
\begin{center}
    $P_d^{OB}= 100*\frac{\sum_{n=1}^{N} (y_{pred}^n==1 \, \&  \, y_{true}^n==1)   }{\sum_{n=1}^{N} (y_{true}^n==1) }$
\end{center}
Here, $N$ is the total number of frequency bands in the digitized spectrum, $y_{pred}$ and $y_{true}$ are the predicted and ground-truth band status, respectively (i.e.,, $0$ for vacant and $1$ for occupied band). It is worth noting that \pdob is the preferred metric for a sparse spectrum consisting of fewer occupied bands since \pdab may give a high value even when the algorithm erroneously detects all bands as vacant. Similarly, \pdab is the preferred metric for spectrum with lower sparsity comprising the higher number of occupied bands. For such spectrum, \pdob may be high even if the algorithm erroneously detects all bands as occupied. We do not consider the spectrum recovery error as a performance metric since it does not offer additional insights for WSS. There are challenges in capturing a large number of samples from hardware for accurate error calculation of spectrum recovery error in real-time. 
\subsubsection{\textbf{Dataset}}
Since SNS-based WSS is based on the underlining assumption of the sparse spectrum, we have generated various datasets with sparsity ranging from 50\%-100\%. We consider the SNS with eight analog-to-digital converters (ADCs), which allow the recovery of the spectrum when the number of occupied bands is less than or equal to 8. We consider two groups of datasets:
\begin{itemize}
    \item Extremely Sparse Spectrum (ESS):  This dataset contains the spectrum with the number of occupied bands between 1 and 3. 
    \item Highly Sparse Spectrum (HSS): This dataset contains the spectrum with the number of occupied bands between 4 and 7.
\end{itemize}

As discussed before, we use \pdob and \pdab as performance metrics for ESS and HSS datasets, respectively.  {\color{black} Interested readers can refer to \cite{codes} for datasets and source code used for the dataset generation.}

\subsubsection{\textbf{OMP without Sparsity Knowledge}}
\begin{figure}[!h]
    \centering
    \includegraphics[width=\linewidth]{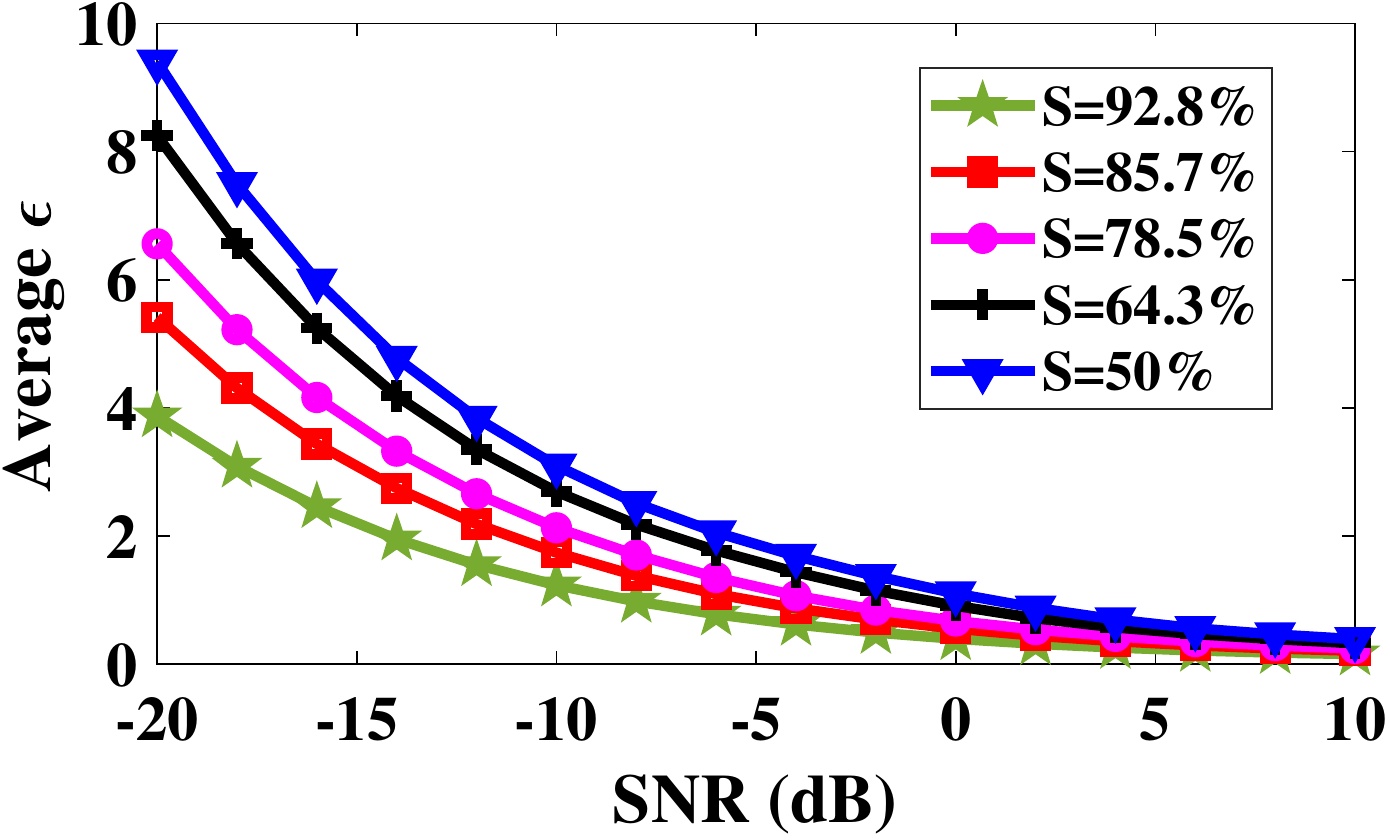}
    \caption{Variation in $\epsilon$ estimates for wide range of spectrum sparsity and SNRs.}
    \label{Fig:fig_1_eps_snr_omp}
\end{figure}

As discussed in Algorithm~\ref{alg:cap}, the number of iterations in OMP algorithm depends on the sparsity, $S$ (Line 5) and this requirement limits its usefulness in realistic applications. We consider the variation of OMP where sparsity knowledge is not known by exploring the stopping criteria for residual, such as $||Res|| <\epsilon$ where $\epsilon$ is the convergence constant. The convergence constant, $\epsilon$, depends on both the sparsity and SNR of the digitized spectrum. To fix $\epsilon$, we assume prior knowledge of the SNR, which is common in wireless systems since wireless receivers can easily measure the SNR of the digitized spectrum. Next, via empirical analysis, we study the correlation between SNR and residual, $||Res||$, to get the desired performance. We use this value of $||Res||$ as an appropriate estimate for $\epsilon$ for a given SNR. In Fig.~\ref{Fig:fig_1_eps_snr_omp}, we compare the desired values of $\epsilon$ for wide range of spectrum sparsity and SNRs. Note that we can not have fixed $\epsilon$ for all SNRs. However, variation in $\epsilon$ for different spectrum sparsity is insignificant, and we can use the average value of $\epsilon$ for all spectrums with varying sparsity. We refer to this algorithm as OMP-$\epsilon$.

\begin{figure}[!h]
    \centering
    \includegraphics[width=\linewidth]{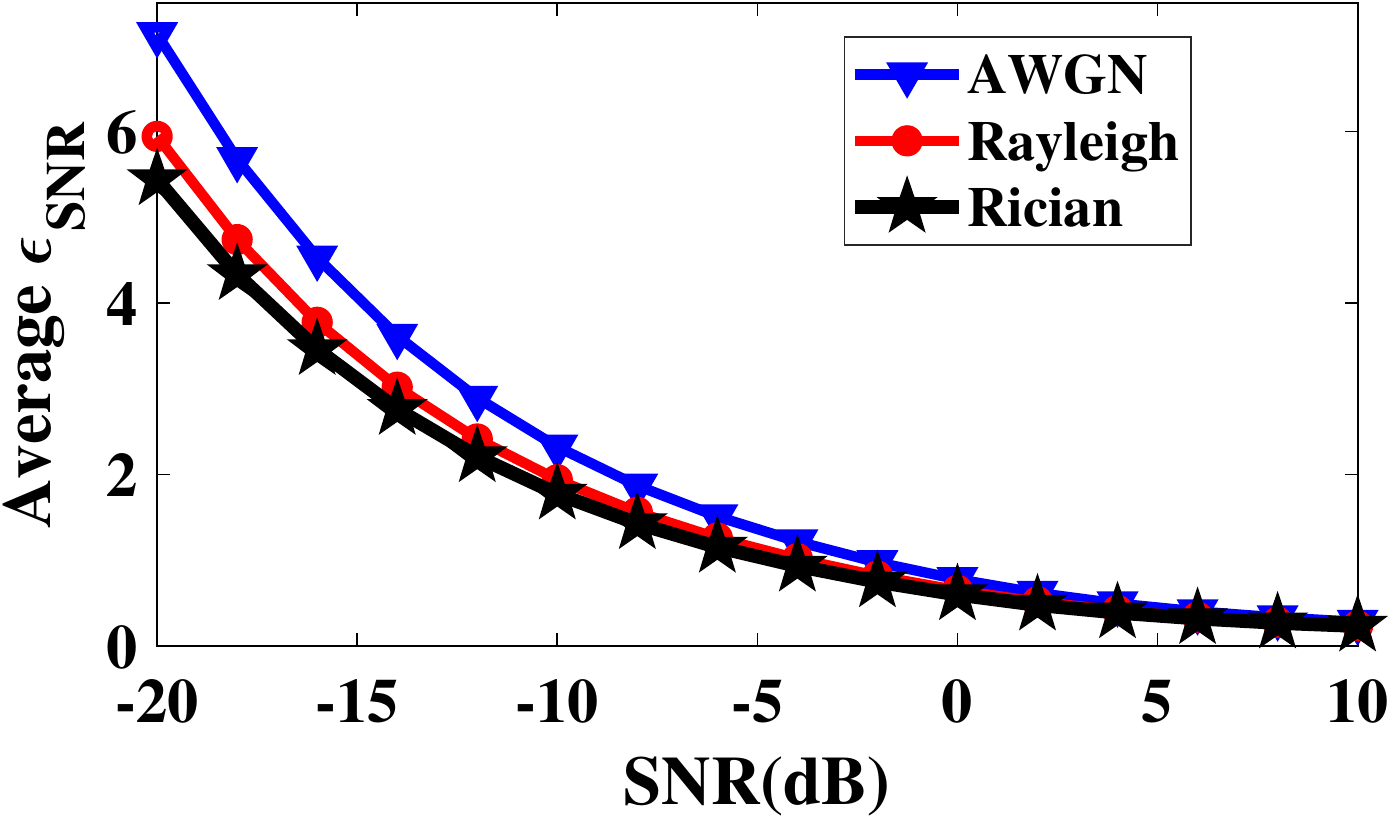}
    \caption{Average $\epsilon_{SNR}$ estimates for various types of wireless channels.}
       \label{Fig:fig_2_eps_snr_omp_chann}
\end{figure}

In a wireless environment, the wireless channel may vary depending on the deployment scenario. In Fig.~\ref{Fig:fig_2_eps_snr_omp_chann}, we study the effect of wireless channel on the average value of $\epsilon$ obtained over wide range of SNRs, $\epsilon_{SNR}$. As expected, the effect of the channel on the WSS is limited since we do not need to recover the spectrum. Thus, we can fix a single value of $\epsilon_{SNR}$ for all types of wireless channels.  

\subsubsection{\textbf{Performance Analysis and Drawbacks}}
\label{Sec:omp_limitations}
\begin{figure}[!t]
    \centering
    \includegraphics[width=\linewidth]{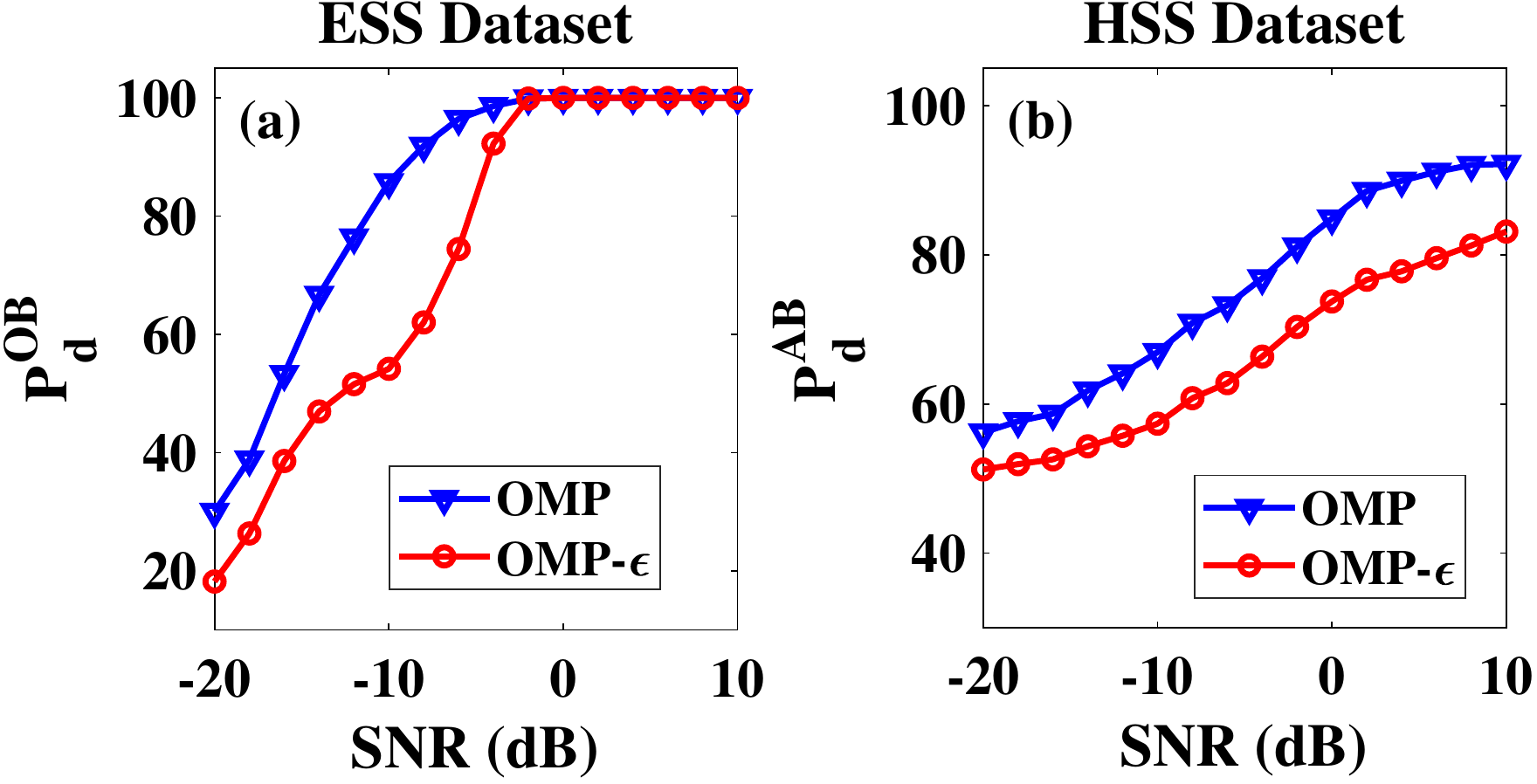}
    \caption{Performance comparisons of the OMP and OMP-$\epsilon$ for (a) ESS and (b) HSS datasets for a wide range of SNRs.}
   \label{Fig:fig_3_Omp_perf}
\end{figure}

In Fig.~\ref{Fig:fig_3_Omp_perf}, we compare the performance of the OMP and OMP-$\epsilon$ for ESS and HSS datasets for a wide range of SNRs. As expected, the performance improves with the increase in SNR. In the case of ESS with \pdob as a performance metric, both OMPs achieve 100\% accuracy at high SNR due to accurate estimation of $\epsilon$. However, the difference between OMP and OMP-$\epsilon$ is around 10-40\% at low SNR. In the case of HSS  with \pdab as a performance metric, overall accuracy is lower, and OMP-$\epsilon$ can not meet the accuracy of OMP even at high SNR. This is because the accurate selection of $\epsilon_{SNR}$ guarantees correct detection of occupied bands but does not guarantee the correct detection of the status of remaining bands. These results indicate that the state-of-the-art OMP algorithm does not offer reliable performance for HSS even at high SNR. 

OMP suffers from multiple drawbacks: 1)~Prior knowledge of sparsity is needed. If such knowledge is unavailable, it suffers from significant performance degradation even with prior knowledge of SNR, 2)~Poor performance in low SNR for both types of spectrum, and 3)~Poor performance even at high SNR for HSS. 
These shortcomings motivates the search for potential alternatives for SNS based WSS.


\section{DLWSS: Spectrum Recovery via Deep Learning  for SNS based WSS}
\label{Sec:DLSWSS_alg}
Previous efforts that aim to tackle SNS-based WSS using deep networks can be broadly classified into two categories: 1)~Iterative approaches \cite{palangi_stacking,Palangi2017ConvolutionalDS,7457684} that follow the OMP formulation and augment the frequency band status detection with a deep network, and 2)~Non-iterative approaches which utilize a deep network for end-to-end spectrum recovery and frequency band status detection \cite{chandhok2020novel}. Since the iterative approaches follow the algorithmic formulation of OMP, they suffer from the same limitations as that of OMP. In this paper, we focus on a non-iterative approach that can handle the signals of varying sparsity and does not require sparsity information or any manually adjusted convergence constant \cite{chandhok2020novel}. The proposed DLWSS algorithm aims to learn an end-to-end model for WSS. 
DLWSS pipeline comprises two stages: 1)~Pre-processing and 2)~Deep Network Prediction.

\subsubsection{\textbf{Pre-Processing}} \label{signal-pre-processing}
{\color{black}Algorithm \ref{alg:cap2} shows the steps involved in the Pre-processing stage of the DLWSS. The pre-processing step receives the signal captured by the antenna and digitized using the SNS-based analog front-end. It processes and normalizes the digitized signal so the DL architecture can handle it. Specifically, this involves computation of the pseudo-recovered spectrum $\tilde X$  using the sensing matrix and sub-Nyquist samples. Since the complex input signal, $\tilde X$ can not be fed directly to the DL architecture; it is converted to a high dimensional real-valued signal as shown in Algorithm~\ref{alg:cap2}. In the end, the signal is normalized for faster convergence of the training process}.



\begin{algorithm}
\caption{Pre-Processing}\label{alg:cap2}
\begin{algorithmic}[1]
\Require \texttt{Sensing matrix A[K][N], aliased sub-Nyquist samples Y[K][Q]}
\State $A_{sq}$= $A^* \times A$ \Comment{Square Matrix $A_{sq}$}
\State $P \times A_{sq}$ = $L \times D \times U $ \Comment{LU Factorization}
\State $A_{sq}^{-1}$=$U^{-1}\times D^{-1} \times L^{-1} \times P^{-1}$
\State $A^\dagger$=$ A_{sq}^{-1} \times A^*$ \Comment{Pseudo-inverse of $A$}
\State $\tilde X \gets A^\dagger \times Y$
\State $\tilde X_d \gets$ \texttt{Concatenate($\tilde X_{real}, \tilde X_{imag}$)}
\State $X_n \gets$ \texttt{Normalize($\tilde X_d$)}
\State \Return $X_n$

\end{algorithmic}
\end{algorithm}


\subsubsection{\textbf{Deep Network Prediction}}
{\color{black} The deep network block of the DLWSS receives the processed input samples from Pre-processing block and predicts the status (vacant or occupied) of each frequency band of the digitized spectrum. The DLWSS architecture is based on a convolutional neural network (CNN) due to its ability to capture spatial correlation in input signals which is integral for the spectrum sensing task. Further, parameter sharing allows them to operate with fewer parameters, enabling the network to be memory efficient and a good candidate for hardware realization}. Table \ref{table:CNN} shows the architecture of the DL model, which comprises three Convolutional (CV) layers and a single Fully Connected (FC) layer. All intermediate activations are Rectified Linear Units (ReLU), and the activation at the output layer is Sigmoid.

 Similar to the deployment of any DL algorithm, DLWSS design has two phases, training and inference. {\color{black}The training phase minimizes a loss function, which measures the difference between the predicted and actual labels}. Since more than one frequency band can be occupied in a wideband spectrum, we formulate the problem as a multi-label binary classification with binary cross-entropy as the training loss function on final sigmoid outputs. {\color{black}After the training mode, the model weights are frozen, and the CNN model is used in inference mode to find the occupancy status of an unknown signal in real-time.} Similar to \cite{chandhok2020novel}, we train the architecture for a dataset using a machine learning framework (PyTorch in our case) and GPUs. After training, model architecture, weights and parameters are extracted for realization on the ZSoC, followed by inference on the new dataset. 
 

\begin{table}[!t]
\caption{CNN based deep network prediction Architecture}
\label{table:CNN}
\renewcommand{\arraystretch}{1.1}
 \resizebox{0.5\textwidth}{!}{
\begin{tabular}{@{}|c|c|c|c|c|@{}}
\hline
\textbf{Layers} & \textbf{Filters} & \textbf{Kernel size} & \textbf{Input Shape} & \textbf{Output Shape} \\ \hline
CV & 256 & 1x150 & 14x299x2 & 14x150x256 \\ \hline
CV & 128 & 1x100 & 14x150x256 & 14x51x128 \\ \hline
CV & 64 & 1x51 & 14x51x128 & 14x1x64 \\ \hline
Flatten & - & - & 14x1x64 & 896x1 \\ \hline
FC & - & - & 896x1 & 14 \\ \hline
\end{tabular}}
\end{table}

\begin{figure*}[!t]
    \centering
    \includegraphics[width=\linewidth]{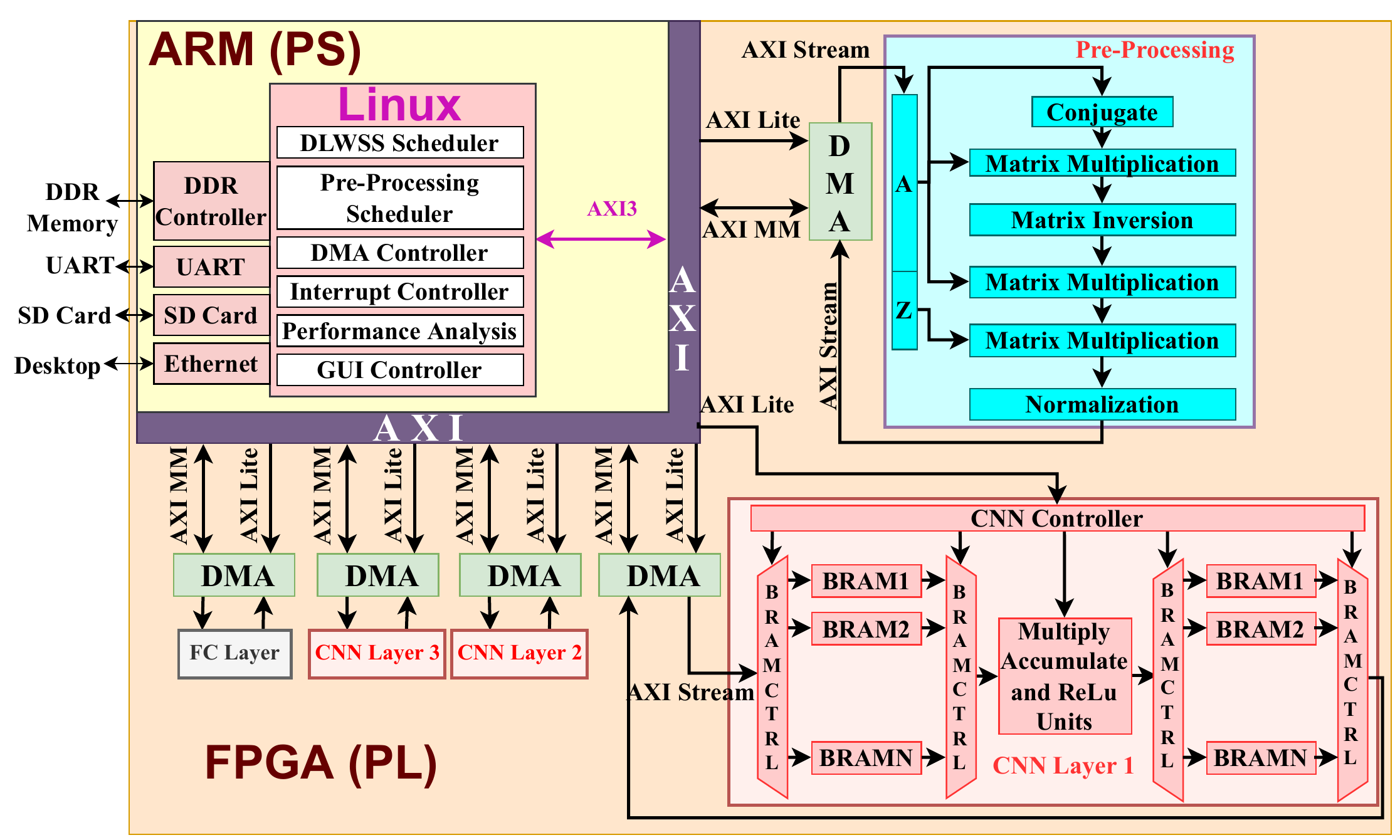}
    \caption{Proposed DLWSS Architecture on Zynq SoC.}
    \label{fig:complete_arch}
\end{figure*}

\section{Realization of DLWSS on ZSoC}
\label{Sec:DLSWSS_arch}
This section presents the complete architecture and mapping of DLWSS on ZSoC, and real-time demonstration via Linux-based graphical user interface (GUI) deployed on ZSoC.
In Fig.~\ref{fig:complete_arch}, various building blocks of the DLWSS such as Pre-processing, CNN and FC layers, scheduler, DMA, interrupt, and GUI controller are shown. For illustration, we have shown the Pre-processing, CNN, and FC layer processing on FPGA. To enable this, we have developed AXI-stream compatible hardware IPs for Pre-processing, CNN, and FC blocks and interconnected them with PS via direct-memory access (DMA) in the scatter-gather mode for efficient data transfers. Later in Section~\ref{Sec:CC}, we considered various architectures via hardware-software co-design by moving the blocks between ARM Processor and FPGA. We have deployed Linux based operating system on PS, which takes care of various scheduling and controlling operations. It also enables GUI development for real-time demonstration.

As discussed in Algorithm~\ref{alg:cap2}, Pre-processing stage involves large-size matrix multiplication and inversion operations. We have modified Xilinx's existing matrix multiplication and matrix inversion reference examples to support the complex number arithmetic since the baseband wireless spectrum is represented using complex samples. The well-known lower-upper (LU) decomposition method is selected for matrix inversion. We have included data buffers using block memory to minimize the repeated data communication between SW and HW and enable matrix operations of different sizes. While the overall algorithm is sequential, we parallelize individual operations like  Matrix Conjugate Transpose and Matrix Multiplication on FPGA. Every element in the matrix is parallelly processed to compute transpose, and every row column dot product in matrix multiplication is performed in parallel to speed up the computation. In addition, multiple instances of these IPs are integrated to get the desired Pre-processing functionality, as shown in Fig.~\ref{fig:complete_arch}. In the end, a normalization block is included to meet the input requirements of a deep neural network.

Next, we focus on mapping each CNN layer on the FPGA. As shown in Fig.~\ref{fig:complete_arch}, CNN involves many multiply and accumulation operations on non-contiguous data, i.e., frequent reading and writing from memory is needed. Depending on the CNN layer dimensions, it may not be possible to store all weights and input data in on-chip memory such as block RAM (BRAM) on FPGA due to limited size and fewer read/write ports. For instance, if we store all the inputs and weights of the CNN model considered in Table~\ref{table:CNN} in the BRAM with single-precision floating (SPFL) number representation, we need a total of 116.4 mega Byte (Mb) of BRAM, assuming the CNN output is written directly in the external DDR memory. Such a large amount of on-chip memory is expensive and may not be available in most SoC. Using external DDR memory leads to frequent data communication overhead resulting in high latency. Thus, mapping the CNN layer on FPGA requires careful sharing of data between external and on-chip memory to get the desired latency. Underlining architecture can be layer-specific depending on the various parameters of the CNN layer, resource, and latency constraints.

\begin{figure}[!h]
    \centering
    \includegraphics[scale=0.3]{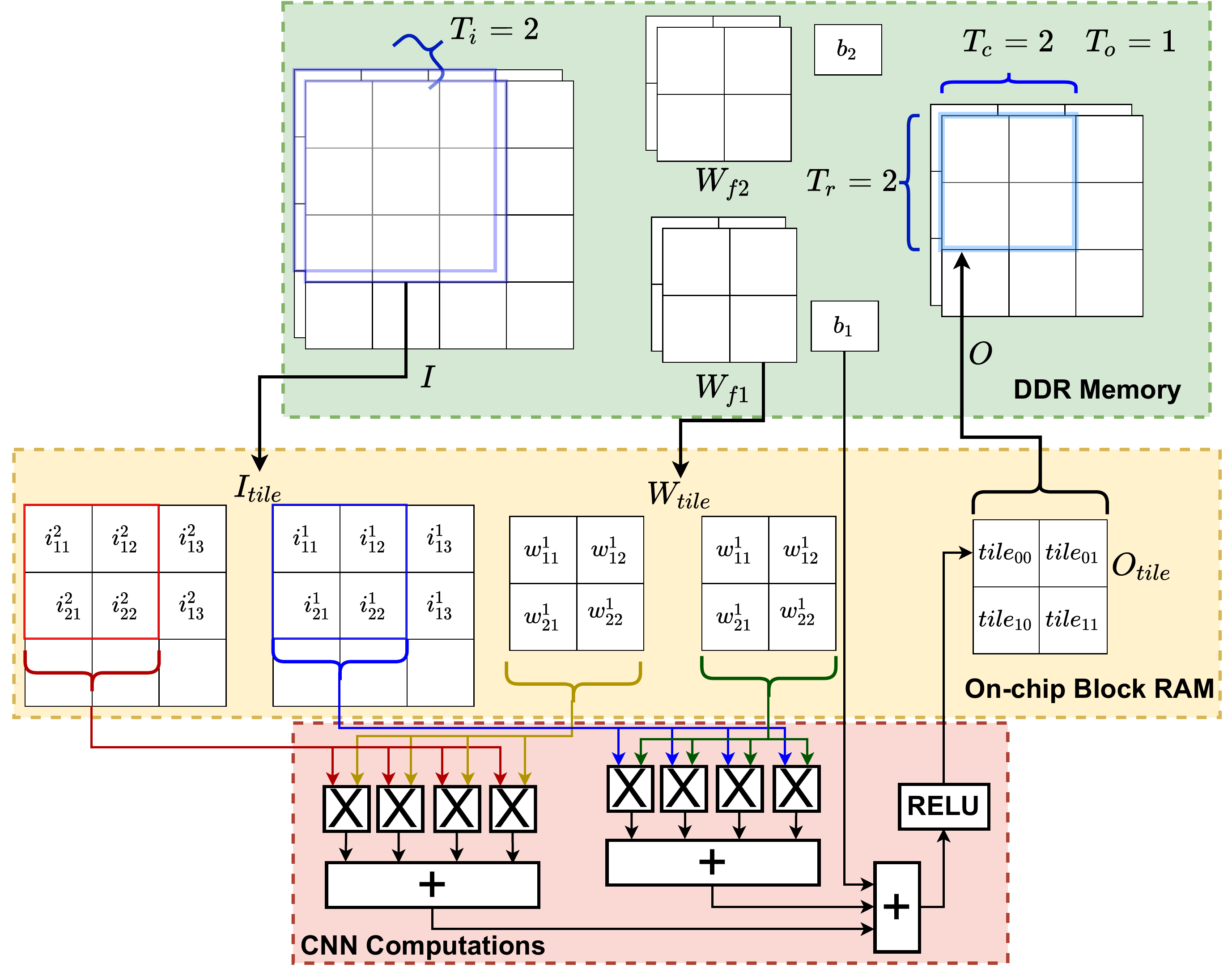}
    \caption{Illustrative example demonstrating the memory tiling approach for realization of the CNN on FPGA.}
    \label{fig:tiling_explanation}
\end{figure}



For the CNN model in Table~\ref{table:CNN}, we have explored the memory tiling approach in which small tiles or blocks of weights and inputs are loaded into the on-chip memory, and care has been taken to maximize the utilization of data currently present in on-chip memory. We define 4 parameters, $<T_o$, $T_i$, $T_r$, $T_c>$ i.e., the tiling factors of output channels, input channels, output rows and output columns, respectively. For easier understanding, we present illustrative example with tiling parameters $<1,2,2,2>$ in Fig.~\ref{fig:tiling_explanation}. The input is of size $4\times4\times2$, i.e, 2 channels and it is convoluted with 2 filters of dimension $2\times2\times2$. Initially, we initialize an output tile and load the input tile and weight tile required to compute the output tile into the on-chip Block RAM. For instance, to compute a $2\times2\times1$ output tile, we need a $3\times3\times2$ input tile ($I_{tile}$) and one filter of dimension $2\times2\times2$ ($W_{tile}$). After loading, the tiled convolution is performed between $I_{tile}$ and $W_{tile}$ using a set of parallel multipliers and adders, referred to as CNN Computations in Fig.~\ref{fig:tiling_explanation}. Once all elements of the output tile are computed, the tile is sent back to the DDR memory, and a similar process is repeated for computing the next output tile. Here, we have shown the computation of one element of the output tile. Depending on resource availability, latency constraints, and memory ports, we can have CNN computations of all elements of a tile or even multiple tiles in a parallel or serial-parallel fashion.

The tiling approach significantly reduces the on-chip memory requirement for the CNN model in Table~\ref{table:CNN}. For instance, tiling with parameters $<20,16,20,20>$ needs only 3.35Mb of BRAM (0.24Mb for output tile,1.464Mb for weight tile and 1.650Mb for input tile) compared to 116.4 Mb in non-tiling based architecture. This in turn allows efficient optimization of hardware IP cores via pipelining and unrolling, resulting in significant improvement in performance. We have explored a wide range of tiling parameters and implemented these architectures on the ZSoC. Please refer to Section~\ref{subsec: tilesize} for more details. We have explored a similar tiling approach for FC layers as well. However, experimental analysis shows that FC layers do not need tiling due to the smaller dimensions of inputs and weights.

\begin{figure}[!h]
    \centering
    \includegraphics[width=\columnwidth]{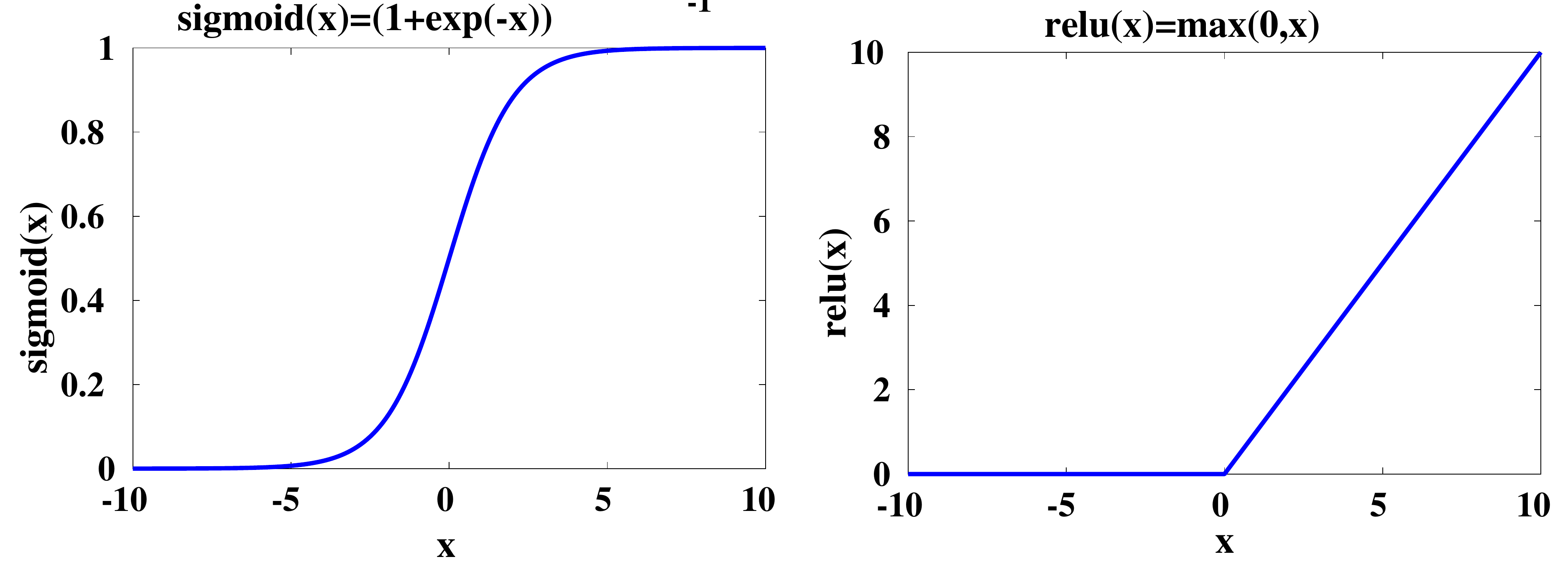}
    \caption{ReLU and Sigmoid activations}
    \label{fig:acti}
\end{figure}

The DLWSS model in Table~\ref{table:CNN} contains two types of non-linear activations; ReLU and Sigmoid. As shown in Fig.~\ref{fig:acti}, the realization of the ReLU on the FPGA is simple as it needs only one comparator and multiplexer. The realization of the Sigmoid on the FPGA can be done either using LUT based approach or by polynomial approximation. The LUT-based approach is memory intensive, while the polynomial-based approach involves many arithmetic and logical operations. We have realized ReLU on FPGA and Sigmoid on ARM processor based on the experimental analysis.


\begin{table*}[!b]
\centering
\caption{HSCD Study of DLWSS Architecture on ZSoC}\label{tab:hscd}
\begin{tabular}{|c|c|c|c|c|c|}
\hline
\rule{0pt}{8pt} \textbf{No.} & \textbf{Blocks on HW} & \textbf{Blocks on SW} & \textbf{Execution Time (s)} & \textbf{\{BRAM, DSP, FF, LUT\}} & \textbf{\{Total Power, Dynamic Power\}(Watts)} \\ \hline
\rule{0pt}{8pt}               1&                                                 & P+CV+FC+A                                                          & 30.8                 & \{0, 0, 0, 0\}                  & \{1.6, 1.2\}                                                                             \\ \hline
\rule{0pt}{8pt} 2 & CV                                                                & P+FC+A                                                            & 2.87                 & \{220, 719, 126741, 100250\}    & \{3.205, 2.205\}                                                               \\ \hline
\rule{0pt}{8pt} 3& CV+FC                                                              & P+A                                                              & 2.869                & \{249, 724, 167676, 117382\}    & \{3.373, 2.369\}                                                               \\ \hline
\rule{0pt}{8pt} 4& P+CV                                                              & FC+A                                                              & 2.863                & \{296, 818, 139857, 117738\}    & \{3.391, 2.387\}                                                               \\ \hline
\rule{0pt}{8pt} 5& P+CV+FC                                                            & A                                                                & 2.863                & \{325, 823, 180792, 134870\}    & \{3.540, 2.533\}                                                               \\ \hline
\rule{0pt}{8pt} 6& P+CV+FC+A                                                          & -                                                               & 2.864                & \{325, 853, 183173, 138230\}    & \{3.567, 2.559\}                                                               \\ \hline
\end{tabular}
\end{table*}

\begin{figure}[!t]
    \centering
    \includegraphics[width=\columnwidth]{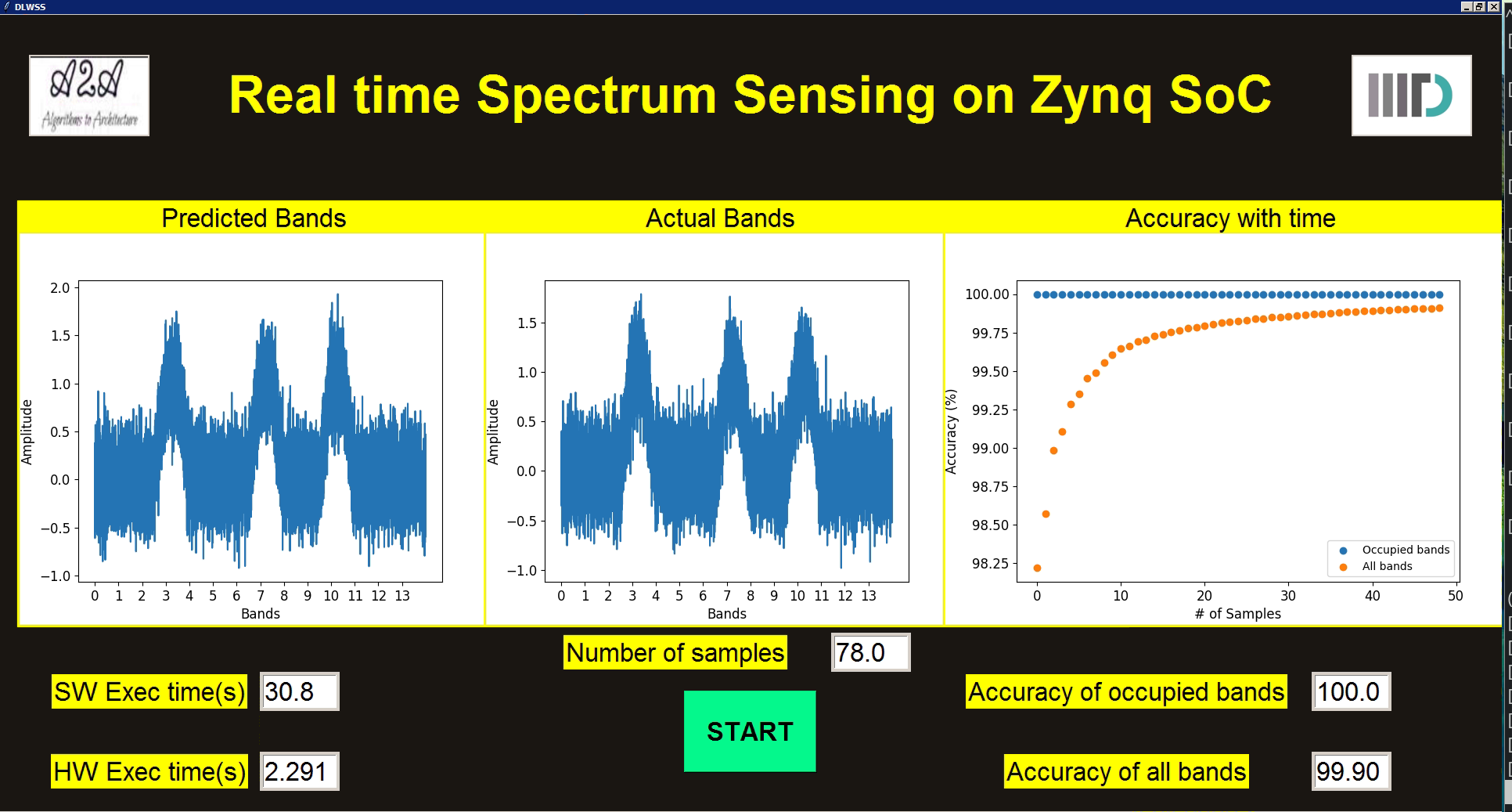}
    \caption{GUI for real-time visualisation of SNS based WSS on ZSoC platform.}
    \label{fig:gui_pic}
\end{figure}

An end-to-end application with a live graphical user interface (GUI) is developed to demonstrate the real-time WSS on the ZSoC using the proposed architectures. The application running on a Petalinux-based operating system deployed on the ARM processor accelerates the computation of pre-processing algorithm and the DL model on the FPGA. The real-time predictions are obtained and stored on the SD-card by the application. The GUI application, shown in  Fig.~\ref{fig:gui_pic}, is deployed on the remote server and reads the contents of SD-card at regular intervals. The GUI is developed using the Tkinter\cite{tkintercite} framework and provides visualization in real-time as and when the architecture predicts the status of frequency bands in the received digitized spectrum. 

\section{Performance Analysis and Comparison with OMP}
\label{Sec:PM}
In this section, we compare the functionality of the OMP and DLWSS architectures for different wireless channels, a wide range of SNRs, and sparsity levels. We consider the effect of prior knowledge of sparsity on performance. As discussed in Section~\ref{section:func_perf_analysis_OMP}, we use \pdob and \pdab as the performance metrics for ESS and HSS spectrums, respectively. For all the experimental results presented in this section, we consider the floating-point arithmetic based architecture realized on ZSoC platform.

\begin{figure}[!t]
    \centering
    \includegraphics[width=\columnwidth]{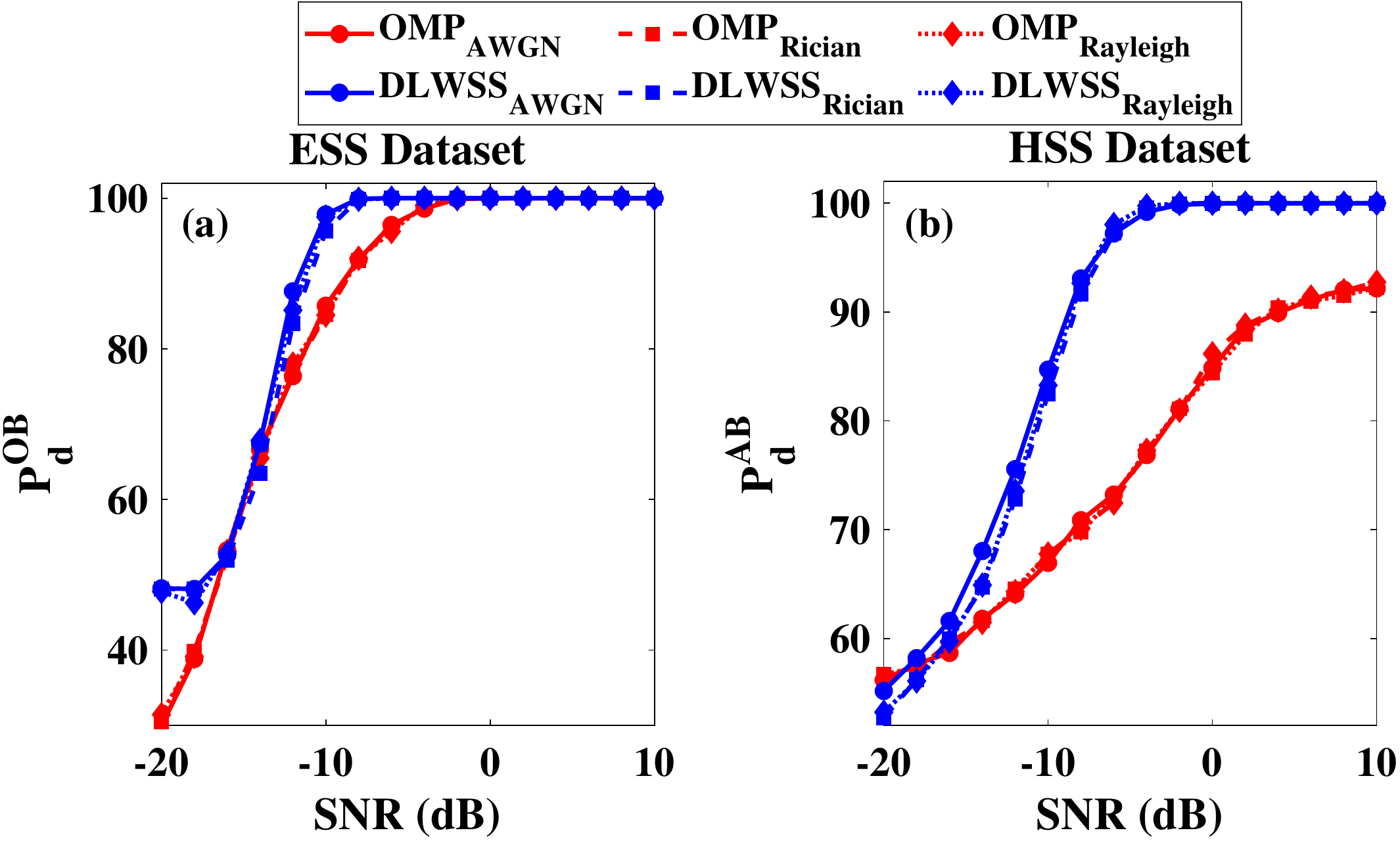}
    \caption{Performance comparison of DLWSS and OMP on (a) ESS and (b) HSS spectrum for different channel conditions. Sparsity level is known.}
    \label{fig:dlwss_float_acc_sparse}
\end{figure}

In Fig.~\ref{fig:dlwss_float_acc_sparse}, we compare the performance of OMP and DLWSS for three different channels (AWGN, Rayleigh, and Rician). We assume the prior knowledge of the spectrum sparsity in the case of OMP. For the ESS spectrum (Fig. \ref{fig:dlwss_float_acc_sparse} (a)), the DLWSS performs better than OMP  with an average performance gain of around 12 \%. We also observed that DLWSS and OMP are robust to changes in channel conditions, which is expected since channel conditions' impact on the spectrum sensing is fairly limited. For the HSS spectrum (Fig.~\ref{fig:dlwss_float_acc_sparse}(b)), the DLWSS significantly outperforms the OMP algorithm with an average performance improvement of around $24\%$. 



\begin{figure}[!t]
    \centering
    \includegraphics[width=\linewidth]{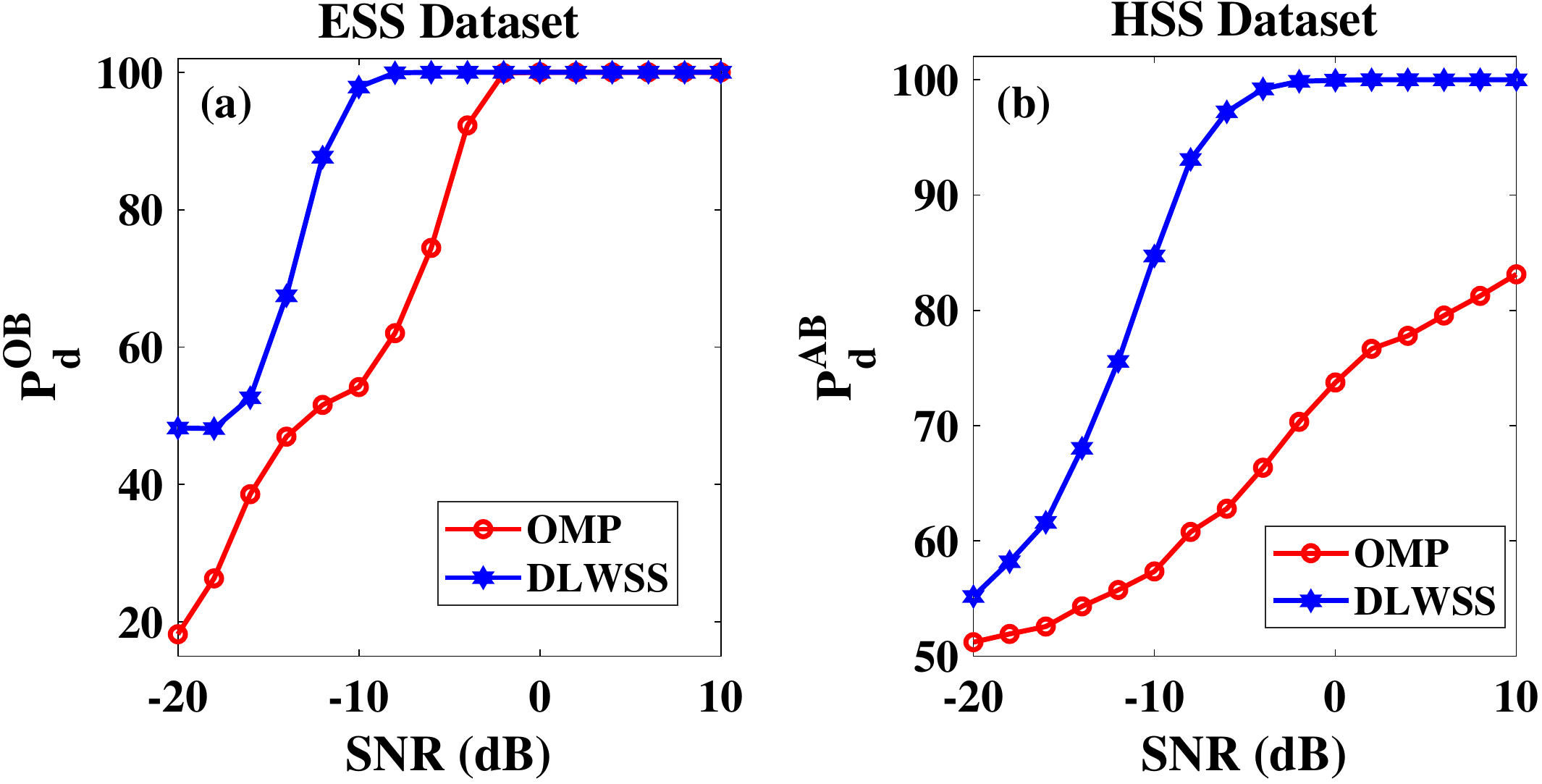}
    \caption{Performance comparison of DLWSS and OMP on (a) ESS and (b) HSS spectrum for different channel conditions. Sparsity level is not known.}
    \label{fig:dense_no_knowledge}
\end{figure}

In practical deployments, spectrum occupancy changes dynamically with time and hence, knowledge of spectrum sparsity is not available. In Fig.~\ref{fig:dense_no_knowledge}, we analyze the performance of the DLWSS and OMP algorithms when spectrum sparsity is not known. It is observed that the proposed DLWSS significantly outperforms the OMP algorithm at all range of SNRs  with an average improvement of around 36.4\% and 31.8\% for for ESS and HSS dataset, respectively. 

\begin{figure}[!t]
    \centering
    \includegraphics[width=\linewidth]{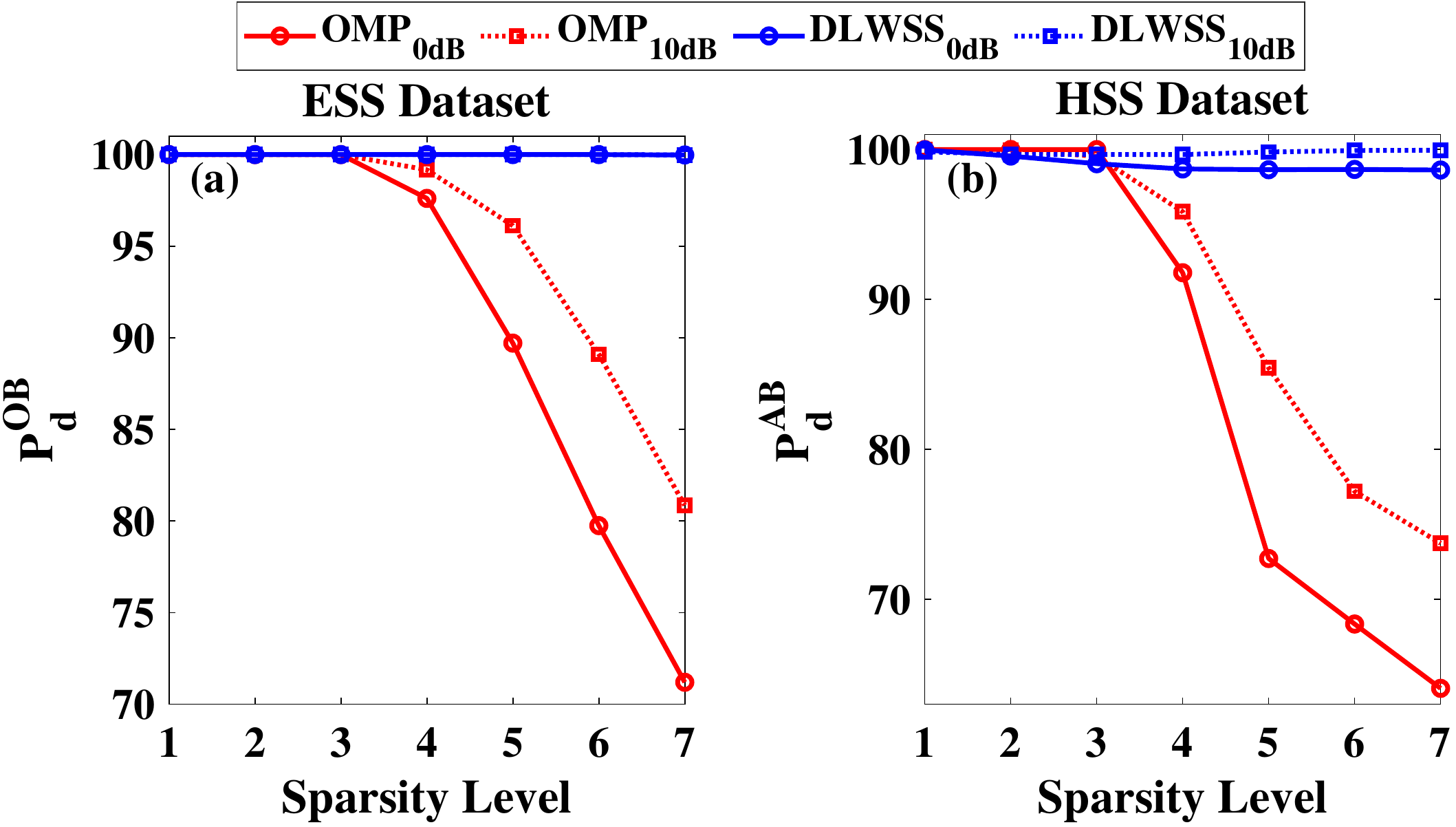}
    \caption{Impact of sparsity level on the performance of the DLWSS and OMP algorithms. }
    \label{fig:individual_sparsity}
\end{figure}

In Fig~\ref{fig:individual_sparsity}, we compare the impact of sparsity on the performance of DLWSS and OMP algorithms when spectrum sparsity is unknown. We consider two SNRs: 0 dB and 10 dB. We trained the DLWSS model using the dataset comprising an equal number of samples from each sparsity level. It can be observed that the DLWSS offers superior performance and significantly outperforms OMP as the sparsity level increases.  Thus, results in Fig.~\ref{fig:dlwss_float_acc_sparse}, Fig.~\ref{fig:dense_no_knowledge} and Fig.~\ref{fig:individual_sparsity} confirms the superiority of the DLWSS over the OMP based approach.

\section{Complexity Analysis}
\label{Sec:CC}
In this section, we analyze the complexity of the DLWSS algorithms for different architectures realized via hardware-software co-design, various word lengths, and memory tiling approaches. 

\subsection{Hardware Software Co-Design (HSCD)}
 A heterogeneous SoC such as Zynq SoC from Xilinx contains ARM Cores as the processing system (SW) and FPGA as the programmable logic (HW). An integral aspect of developing an efficient mapping of the DLWSS on the ZSoC is to design an optimal HSCD strategy to facilitate functionally accurate architecture for the desired latency and resource constraints. Specifically, we need to decide how to partition the algorithm between HW and SW and minimize the data communication overhead between them. The HSCD is important since sequential operations, scheduling tasks, and GUI are preferred on SW. At the same time, FPGA can efficiently handle the task, which can be accelerated via parallel processing. However, in certain situations, serial tasks are preferred on HW, while parallel tasks are preferred on SW to avoid data communication overhead between SW and HW.
Furthermore, some operations may offer speed up on HW. Still, such speed up may not be significant compared to the algorithm's overall execution time; hence, realizing such tasks on SW can reduce FPGA size, cost, and power consumption. Such trade-offs demand a detailed study of various HSCD architectures to design an architecture that meets the cost, latency, and power constraints. 
 



Table~\ref{tab:hscd} shows the results of our HSCD study for the DLWSS architecture comprising pre-processing (P), convolution (CV), fully connected (FC), and activations (A) blocks. 
As shown in Row 1, mapping the entire architecture on SW results in a high execution time of 30.8 seconds ($s$). As we shift more blocks to HW, the execution time is reduced while resource and power consumption increase (Rows 2-6). The CV layer is the most computational complex unit in the DLWSS, and realizing it on the HW results in a significant reduction in the execution time from 30.8$s$ to 2.87$s$.

Since the DLWSS model has fewer FC layers as compared to the CV layers, shifting the FC layer to HW does not offer substantial improvement in execution time compared to 2.7\%, 0.6\%, 9.4\% and 7.8\% increase in BRAM, DSP, FF and LUT utilization, respectively. Similarly, shifting the pre-processing algorithm (P) on HW leads to improvement in execution time of around 7 $ms$ compared to increase in the BRAM, DSP and LUT usage by 7\%, 11\% and 8\%, respectively with respect to previous architecture in Row 3. 
As we move FC block to HW (Row 5), we notice that the impact of FC block is the same as seen in Row 2, which is expected as adding the FC block adds a constant time and constant resource utilization for all cases. 
In Row 6, we move Sigmoid based activation block to HW which results in increase in the DSP and LUT utilization by 3\% and 1.5\%, respectively, with around 1 $ms$ improvement in execution time. Note that as we gradually shift  more and more blocks from SW to HW, the overall power consumption increases from 3.205W (Row 2) to 3.567W (Row 6).
Thus, from HSCD perspective, it is better to keep all blocks except CV layers on the SW given high resource penalty in HW for small gain in execution time.



\subsection{Word Length Optimization}

Conventionally, HW realization of the algorithm in floating-point arithmetic offers good functional accuracy but incurs high resource utilization, power consumption, and execution time. Since the extremely-large dynamic range offered by floating-point arithmetic may not be needed for all the sub-blocks of the algorithm, fixed-point arithmetic can potentially offer a significant reduction in resource, power, and execution time without compromising on the functional accuracy. In wireless applications such as DLWSS, the dynamic range of inputs, weights, and activation is limited due to analog-to-digital converters (ADCs); hence, fixed-point architectures are preferred. In this section, we discuss the selection of appropriate word length for the part of the algorithm realized on the HW, i.e., FPGA, and its impact on the functional accuracy, resource utilization, execution time, and power consumption.


We use $W$ bits to represent each number in fixed point quantization. Out of $W$ bits, we use  $I$ bits to represent the integer part and ($W-I$) bits to represent the fractional part. For example, the fixed-point representation of a number, 3.25, needs only 6 bits with $I=3$ compared to 64 bits in double-precision floating point (DPFP), 32 bits in single-precision floating point (SPFP), and 16 bits in half-precision floating-point (HPFP). Thus, depending on the dynamic range of the given variable, the appropriate selection of $W$ and $I$ can avoid loss in functional accuracy. 
To identify appropriate values of $W$ and $I$ for the DLWSS architecture, we analyzed the dynamic range of inputs, outputs, and intermediate outputs of various sub-blocks. For instance, Table~\ref{table:fp_comp} shows the analysis to determine the optimal integer width for hardware realization of CV and FC layers of the DLWSS, where we infer the model on samples of the dataset to estimate the ranges of intermediate activations and the model weights. It can be observed that the minimum value of $I$ is 9 and 2 bits for activation and weights, respectively. The value of $W$ depends on the number of bits for accurate fractional number representation to get the desired functional performance, and we select them via heuristic experiments.


\begin{table}[!h]
\centering
\caption{Integer Word Length Selection for DLWSS}
\begin{tabular}{|c| c| c| c| c|} 
 \hline
\rule{0pt}{7pt} \textbf{Type} & \textbf{Layer} & \textbf{Minimum Value} & \textbf{Maximum Value} & $I_{min}$\\ [0.5ex] 
 \hline
\rule{0pt}{7pt} Activation & CV & -77.061 & 199.309 & 9  \\ 
 \hline
\rule{0pt}{7pt} Activation & FC  &  -86.594 & 158.975 & 9\\ 
 \hline
\rule{0pt}{7pt} Weight & CV & -0.4812 & 0.9561 & 2  \\ 
 \hline
\rule{0pt}{7pt} Weight & FC  &  -0.0661 & 0.0219 & 2\\ 
 \hline
\end {tabular}

\label{table:fp_comp}
\end{table}

\begin{table*}[!t]\centering
\caption{Functionality and Complexity Analysis of DLWSS Architectures for Different Word Lengths at 10 dB SNR}
\label{tab:word_length_table}
\renewcommand{\arraystretch}{1.2}
 \resizebox{\textwidth}{!}{
\begin{tabular}{@{}|c|c|c|c|c|c|c|c|@{}}
\hline
\textbf{No. }&  \begin{tabular}[c]{@{}c@{}}\textbf{Activation}\\\textbf{WL:} \textless{}$W_a, I_a$\textgreater{}\end{tabular}   & \begin{tabular}[c]{@{}c@{}}\textbf{Weights}\\\textbf{WL:} \textless{}$W_w, I_w$\textgreater{}\end{tabular} & \textbf{Execution Time (s)} & \textbf{ESS:} $P_d^{OB}$ & \textbf{HSS:} $P_d^{AB}$ & \textbf{\{BRAM, DSP, FF, LUT\} }&\textbf{ \{Tot. Power, Dyn. Power\}} \\ \hline
1& SPFP & SPFP & 2.28 & 100 & 100 & \{256,879,150107,117023\} & \{4.218,3.944\} \\ \hline
2& HPFP & HPFP & 2.11 & 100 &  99.97 & \{213,719,98614,74952\} & \{2.688,2.448\} \\ \hline

3& \textless{}29,9\textgreater{} & \textless{}16,2\textgreater{} & 2.28 & 100 &  100 & \{248,399,75935,81119\} & \{2.523,2.285\} \\ \hline
4& \textless{}28,9\textgreater{} & \textless{}16,2\textgreater{} & 2.28 & 100 &  100 & \{248,399,71907,74381\} & \{2.484,2.247\} \\ \hline
5& \textless{}27,9\textgreater{} & \textless{}16,2\textgreater{} & 2.30 & 100& 99.98 & \{243,399,59641,50259\} & \{2.353,2.119\} \\ \hline
6& \textless{}26,9\textgreater{} & \textless{}16,2\textgreater{} & 2.30 & 100 &  99.94 & \{240,399,59641,50259\} & \{2.351,2.114\} \\ \hline
7&\textless{}25,9\textgreater{} & \textless{}16,2\textgreater{} & 2.30 & 100 &  99.94 & \{240,399,53926,41443\} & \{2.299,2.065\} \\ \hline
8&\textless{}24,9\textgreater{} & \textless{}16,2\textgreater{} & 2.30 & 100 &  98.78 & \{232,399,53926,41443\} & \{2.294,2.061\} \\ \hline
9&\textless{}23,9\textgreater{} & \textless{}16,2\textgreater{} & 2.28 & 87.38 &  79.45 & \{232,399,53926,41443\} & \{2.294,2.061\} \\ \hline
10&\textless{}22,9\textgreater{} & \textless{}16,2\textgreater{} & 2.28 & 70.96 &  57.46 & \{232,399,53926,41443\} & \{2.294,2.061\} \\\hline

\end{tabular}}
\end{table*}

We have designed and implemented DLWSS architectures of various WLs on ZSoC and analyzed their performance to identify appropriate WL for fractional number representation. In Table~\ref{tab:word_length_table}, we compare 10 different DLWSS architectures with fixed integer WL of 9 for activation and 2 for weights. Here, we have fixed the WL of weights to \textless{}16,2\textgreater{} to identify the WL for activation.  A similar process is done to identify the WL of weights by fixing the WL of activation. Corresponding details are omitted to avoid repetition of discussion. All these results are evaluated for a fixed tile size of \textless{}20,16,20,20\textgreater{}. Please refer to Section~\ref{subsec: tilesize} for more details about impact of tile size on resource utilization.

To begin with, we consider DLWSS with SPFL WL in Row 1 of Table~\ref{tab:word_length_table}. As expected, it offers excellent functional accuracy in terms of the chosen performance metrics. In Row 2, we consider DLWSS with HPFP WL. It offers nearly identical functional performance as that of the SPFP architecture with savings of around ${7\%}$ in BRAM, 17\% in DSP, and 46 in \% LUT and power savings of ${0.47W}$. Also, there is a significant reduction of 0.17$s$ in execution time. 

Next, we have explored eight different DLWSS architectures via fixed-point quantization, and corresponding results are given in Rows 3-10 of Table~\ref{tab:word_length_table}. As expected, there is a slight degradation in functional accuracy as WL decreases. The DLWSS architecture with the WL of \textless{}23,9\textgreater{} or below in Rows 9-10 suffers from significant degradation in performance and should be avoided. The DLWSS architecture in Row 7 with a WL of 25 offers functional performance same as that of floating-point architectures with more than 50\% savings in DSP, FFs, and LUTs over HPFP architecture in Row 2 and over 60\% savings in DSP, FFs, and LUTs over SPFP architecture in Row 1. These savings can be further improved using the DLWSS architecture with WL of \textless{}24,9\textgreater{} in Row 8 with minor degradation in performance. Note that we have assumed that the inputs and outputs are in SPFP format; hence, additional WL conversion inside the architecture is needed. We can reduce the execution time further if the input and output WLs are the same as the rest of the architectures.

\begin{figure}[!h]
    \centering
    \includegraphics[width=\linewidth]{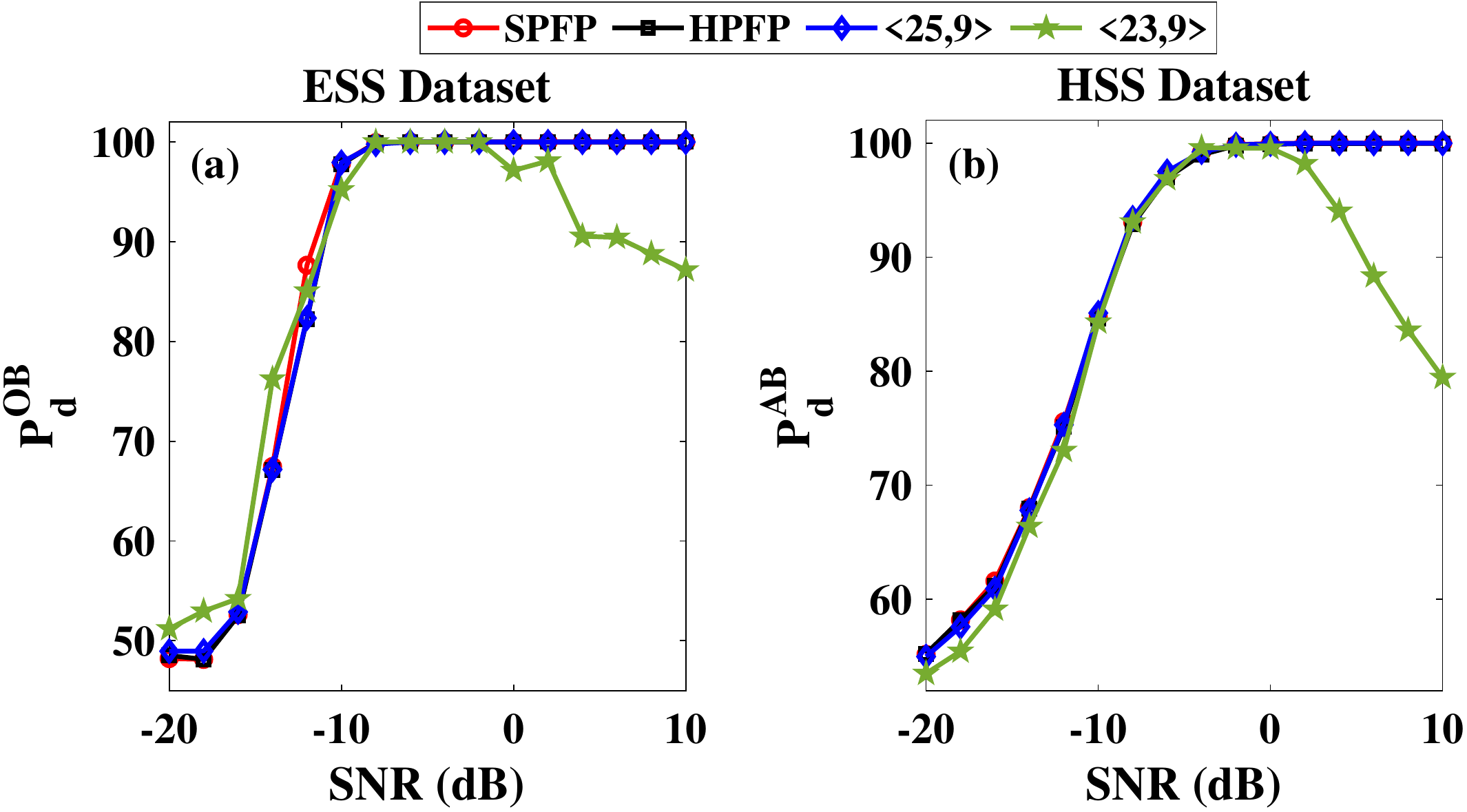}
    \caption{Impact of the WL on the functional accuracy of the DLWSS architecture.}
    \label{fig:WLperf}
\end{figure}

In Table~\ref{tab:word_length_table}, we have fixed the SNR to 10 dB. To analyze the architecture's performance at different SNRs, we have compared the \pdob for ESS and \pdab for HSS for SNRs ranging from -20dB to 10dB in Fig.~\ref{fig:WLperf}. It can be observed that the performance degrades at higher SNR due to insufficient WL.

\subsection{Impact of Memory Tiling Approach} \label{subsec: tilesize}
\begin{figure}
    \centering
    \includegraphics[width=\columnwidth]{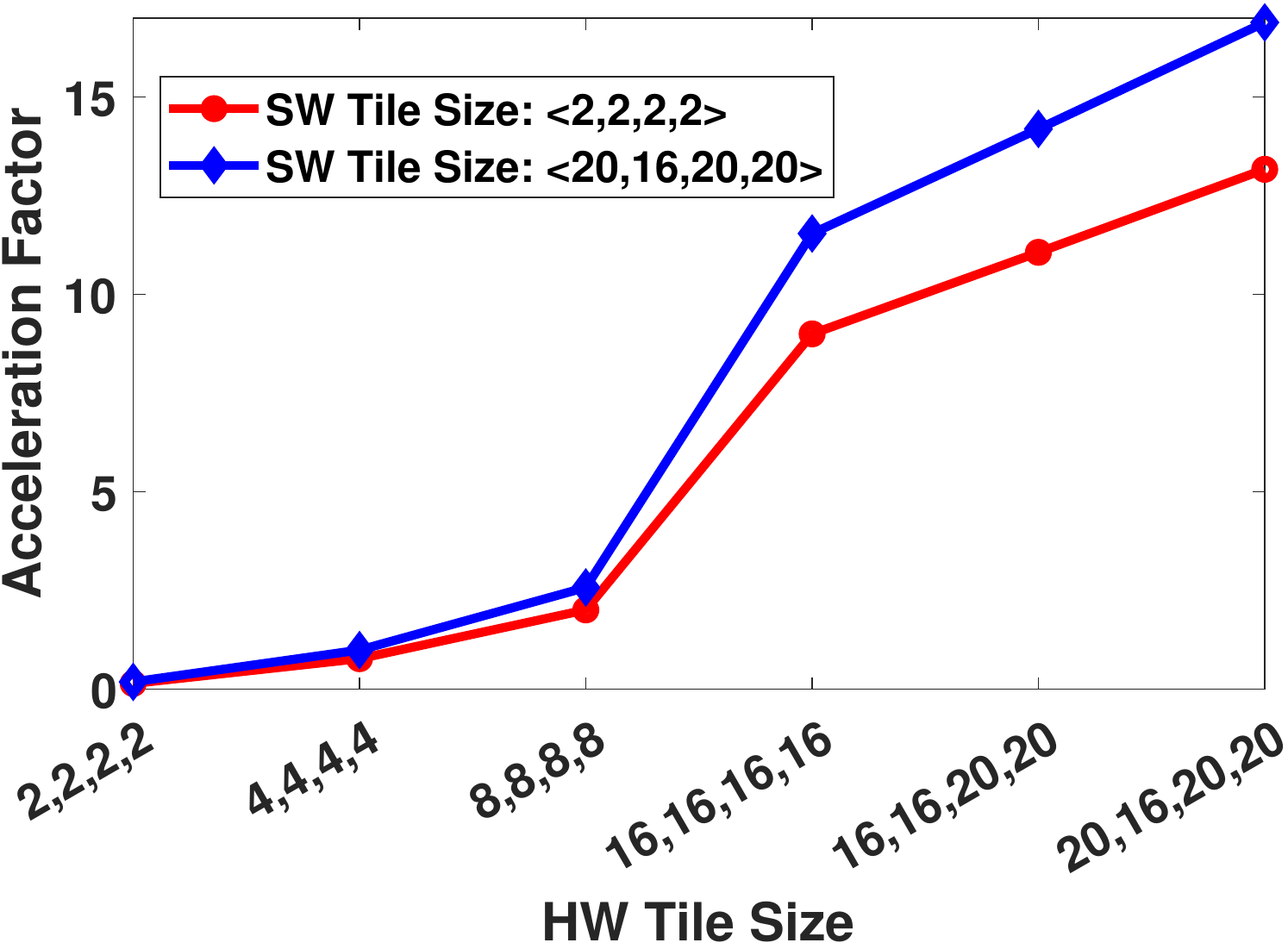}
    \caption{Impact of tile size on the execution time acceleration factor.}
    \label{fig:tilingimpactontime}
\end{figure}

As discussed in Section~\ref{Sec:DLSWSS_arch}, memory tiling is essential to enable an efficient implementation of the memory-intensive convolution layer operations on HW. In this section, we discuss the impact of tiling size on the performance of the DLWSS architecture realized on the SW and HW. In Fig.~\ref{fig:tilingimpactontime}, we compare the acceleration factor, i.e. the ratio of execution time on SW and HW, for different tiling sizes on SW and HW realizations. It can be observed that the acceleration factor increases with the increase in tiling size of the HW architecture for a given tiling size of SW realization. Such behavior can be attributed to the following: 
\begin{enumerate}
    \item Higher tile size on HW means fewer accesses to external DDR memory since more data is buffered in on-chip block RAM. 
    \item Higher tile size allows more opportunities for intra-tile parallelization i.e. dot products within a tile can be computed in parallel on HW. 
\end{enumerate}

Interestingly, an increase in the tiling size of the SW realization results in degradation in the execution time of the DLWSS on SW. For instance, acceleration factor is higher for SW tiling size of $<20,16,20,20>$ compared to SW tiling size of $<2,2,2,2>$. This behavior is exactly reverse compared to that of HW, and one possible reason is data caching in SW realization. Smaller tile size allows tiles to be cached in the local data cache memory of ARM cores, thereby limiting the number of accesses to external DDR memory. Since cache size is significantly smaller than on-chip BRAM on HW, larger tile size results in frequent cache flush requirements, resulting in execution time degradation. 

Though the tiling size does not impact functional accuracy, a larger tiling size leads to higher resource utilization, as shown in Table~\ref{tab:tilingresources}. This is because a larger tile size needs a higher amount of on-chip memory and enables parallel computations due to the availability of more data on the chip. The use of fixed-point arithmetic can help to reduce resource utilization significantly. Thus, combining hardware-software co-design, WL, and tiling parameters are vital to meet the given resource and execution time constraints. 


\begin{table}[!h]
\centering
\caption{Effect of Tiling and WL on resource utilization}
\label{tab:tilingresources}
\renewcommand{\arraystretch}{1.3}
\resizebox{\columnwidth}{!}{%
\begin{tabular}{|c|c|ccccc|}
\hline
\multicolumn{1}{|l|}{\multirow{2}{*}{\textbf{Resources}}} & \multicolumn{1}{l|}{\multirow{2}{*}{\textbf{Word-Length}}} & \multicolumn{5}{c|}{\textbf{Memory Tiling Parameters}} \\ \cline{3-7} 
\multicolumn{1}{|l|}{} & \multicolumn{1}{l|}{} & \multicolumn{1}{c|}{\textbf{\textless{}2,2,2,2\textgreater{}}} & \multicolumn{1}{c|}{\textbf{\textless{}8,8,8,8\textgreater{}}} & \multicolumn{1}{c|}{\textbf{\textless{}16,16,16,16\textgreater{}}} & \multicolumn{1}{c|}{\textbf{\textless{}16,16,20,20\textgreater{}}} & \textbf{\textless{}20,16,20,20\textgreater{}} \\ \hline
\multirow{2}{*}{\textbf{BRAM}} & \textbf{Floating Point} & \multicolumn{1}{c|}{14.9} & \multicolumn{1}{c|}{67.9} & \multicolumn{1}{c|}{219.96} & \multicolumn{1}{c|}{219.96} & 255.72 \\ \cline{2-7} 
 & \textbf{Fixed Point} & \multicolumn{1}{c|}{14.5} & \multicolumn{1}{c|}{63.5} & \multicolumn{1}{c|}{203.5} & \multicolumn{1}{c|}{203.5} & 239.5 \\ \hline
\multirow{2}{*}{\textbf{DSP}} & \textbf{Floating Point} & \multicolumn{1}{c|}{82} & \multicolumn{1}{c|}{239} & \multicolumn{1}{c|}{719} & \multicolumn{1}{c|}{719} & 879 \\ \cline{2-7} 
 & \textbf{Fixed Point} & \multicolumn{1}{c|}{81} & \multicolumn{1}{c|}{143} & \multicolumn{1}{c|}{335} & \multicolumn{1}{c|}{335} & 399 \\ \hline
\multirow{2}{*}{\textbf{LUT}} & \textbf{Floating Point} & \multicolumn{1}{c|}{35107} & \multicolumn{1}{c|}{49097} & \multicolumn{1}{c|}{100249} & \multicolumn{1}{c|}{100687} & 117016 \\ \cline{2-7} 
 & \textbf{Fixed Point} & \multicolumn{1}{c|}{35652} & \multicolumn{1}{c|}{37656} & \multicolumn{1}{c|}{46824} & \multicolumn{1}{c|}{47630} & 59259 \\ \hline
\end{tabular}%
}
\end{table}

\section{Conclusions}
\label{Sec:conc}
In this paper, we designed and implemented statistical orthogonal matching pursuit (OMP), and deep learning (DL) based algorithms on Zynq System-on-chip (SoC) for wideband sensing applications. We have provided in-depth experimental results and complexity comparisons among various architectures obtained via hardware-software co-design, word-length optimization, and memory tiling. Specifically, we demonstrated the drawbacks of conventional OMP algorithms, such as poor performance at a low signal-to-noise ratio (SNR) and the need for prior knowledge of sparsity. These drawbacks are addressed via a novel DL-based approach. However, the DL architecture's high resource utilization and execution time is a concern. 
In the future, we also plan to explore Neural architecture search (NAS) along with quantized model training to reduce the complexity of DL architecture. We also plan to integrate the proposed architecture with analog-front-end for experiments with real-radio signals.  

\bibliographystyle{IEEEtran}
\bibliography{biblio.bib}

\end{document}